# Population genomics of transitions to selfing in Brassicaceae model systems

Tiina M. Mattila[1], Benjamin Laenen[2], Tanja Slotte[2]

[1]Department of Ecology and Genetics, University of Oulu, 90014 Oulu, Finland

[2]Department of Ecology, Environment, and Plant Sciences, Science for Life Laboratory, Stockholm University, 10691 Stockholm, Sweden

**Running head:** Population genomics of shifts to selfing

**Abstract**

Many plants harbor complex mechanisms that promote outcrossing and efficient pollen transfer. These include floral adaptations as well as genetic mechanisms, such as molecular self-incompatibility (SI) systems. The maintenance of such systems over long evolutionary timescales suggests that outcrossing is favorable over a broad range of conditions. Conversely, SI has repeatedly been lost, often in association with transitions to self-fertilization (selfing). This transition is favored when the short-term advantages of selfing outweigh the costs, primarily inbreeding depression. The transition to selfing is expected to have major effects on population genetic variation and adaptive potential, as well as on genome evolution. In the Brassicaceae, many studies on the population genetic, gene regulatory and genomic effects of selfing have centered on the model plant *Arabidopsis thaliana* and the crucifer genus *Capsella*. The accumulation of population genomics datasets have allowed detailed investigation of where, when and how the transition to selfing occurred. Future studies will take advantage of the development of population genetics theory on the impact of selfing, especially regarding positive selection. Furthermore, investigation of systems including recent transitions to selfing, mixed mating populations and/or multiple independent replicates of the same transition will facilitate dissecting the effects of mating system variation from processes driven by demography.

**1. Introduction**

Flowering plants harbor a great variety of mating systems and associated floral and reproductive adaptations *(1)*, and there is a rich empirical and theoretical literature on the causes of this diversity *(2-6)*. About half of all flowering plants harbor genetic self-incompatibility (SI), a molecular recognition system that allows plants to recognise and reject self pollen, and that has arisen multiple times in the history of flowering plants *(7)*. Despite the fact that molecular SI systems are widespread, loss of SI, often accompanied by a shift to higher selfing rates, has occurred even more frequently, in many independent plant lineages *(8)*. This transition can be favored under conditions when the benefits of selfing, such as reproductive assurance *(2)* and the 3:2 inherent genetic transmission advantage of selfing *(9)*, outweigh the costs of inbreeding depression and reduced opportunities for outcrossing through pollen (pollen discounting). As the favorability of the transition hinges on ecological factors including access to mates and pollinators that may vary greatly spatially or temporally, it is perhaps not surprising that the transition to selfing has occurred repeatedly *(10)*. Over a longer term, however, the loss of SI is associated with a reduction in the net diversification rate *(11)*, a finding that provides tentative support for Stebbins' suggestion that selfing is an evolutionary dead-end *(12)*. While the underlying ecological and evolutionary mechanisms behind this observation remain unclear, it was suggested already by Stebbins *(12)* that decreased adaptive potential in selfers would lead to higher extinction rates, a suggestion that is supported by theoretical modeling *(13)*. However, selfing does not only affect adaptation but also the impact of purifying selection *(14, 15)*, and the relative importance of accumulation of deleterious mutations vs. reduced potential for adaptation in selfing lineages currently remains unclear. To fully understand the impact of mating system shifts on evolutionary processes, it is necessary to combine theoretical and empirical investigations, and ideally to study several parallel transitions from outcrossing to selfing. Molecular population genetics has proven to be a powerful tool to shed light on the role of natural selection in shaping the patterns of variation in selfing species. Here we will give an outline of recent work in this area, with a focus on two main model systems in the Brassicaceae.

**2. The molecular basis of the loss of SI and evolution of self-fertilization in Brassicaceae**

The effects of the transition to self-fertilization on population genomic variation and molecular evolution have been extensively studied in two systems from the Brassicaceae family, *Arabidopsis* and *Capsella*. Both of these genera have outcrossing SI as well as SC species with high selfing rates, and thus serve as good models to study this evolutionary transition *(16-18)*. The most widely studied SC species are *Arabidopsis thaliana* and *Capsella rubella*, which have both been estimated to be highly selfing *(19-22)* The patterns of variation and molecular evolution in these selfers are often contrasted with those in their diploid sister species *Arabidopsis lyrata* and *Capsella grandiflora*, which are both SI and outcrossing. Investigations of the other SC species from these genera such as allopolyploid *Arabidopsis suecica (23)* and *Arabidopsis kamchatica (24-26)*, diploid *Capsella orientalis* and allopolyploid *Capsella bursa-pastoris (18)* have given further insight into the evolution of selfing. In Figure 1, we provide an overview of the evolutionary relationships among the best-studied *Arabidopsis* and *Capsella* species.

Knowledge on the molecular basis of the breakdown of SI is at the center of studies investigating the early genetic causes of the transition to selfing in the Brassicaceae. In Brassicaceae, the SI recognition system includes two key genes at the non-recombining self-incompatibility locus (*S*- locus) as well as modifier genes. The gene *SRK* encodes a *S*-locus receptor kinase that is located on the stigma surface and acts as the female specificity determinant, whereas the gene *SCR* encodes a pollen ligand that is deposited on the pollen surface and acts as the male specificity determinant *(27)*. This reaction is a key-lock protein interaction between the female determinant on the stigma and the male determinant on the pollen coat *(27)*. When SRK on the stigma binds to SCR from the same *S*-haplotype, a downstream reaction is triggered which culminates in the prevention of pollen tube growth and fertilization *(28)*. The evolution of selfing proceeds by disruptions of the SI reaction, for example due to loss-of-function mutations in key *S*-locus genes or in unlinked modifier genes (e.g. *(29)*), after which the selection at these loci is relaxed and the genes may be degraded further.

There has been intense interest in the role of parallel molecular changes underlying repeated shifts to selfing associated with the loss of SI (reviewed in *(30)*). In particular, theory predicts that mutations that disrupt the function of the male specificity determinant might spread more easily than those that disrupt the female specificity determinant *(31, 32)*, and there is accumulating support for this prediction. For instance, in *A. thaliana*, Tsuchimatsu et al. (2010) *(33)* showed that an inversion in *SCR* underlies SC in many European accessions. In some accessions, SI can be restored by introduction of functional SRK-SCR allele from self-incompatible sister species *A. lyrata* but variability between accessions exists *(34, 35)*. In *Capsella* homologous machinery has been shown to underlie the SI reaction *(36)* and there is widespread transspecific shared polymorphism between *C. grandiflora* and SI *Arabidopsis* species at the *S*-locus *(37)*. The loss of SI in *C. rubella* is due to changes at the *S*-locus *(36, 38)*, and experiments suggest that breakdown of the male specificity function is responsible for this loss *(36)*. However, the molecular basis of the breakdown of SI in *C. rubella* remains unclear *(39)* and this is also true for the other SC *Capsella* species. With recent progress in long-read sequencing, which facilitates assembly of the *S*-locus, this area is ripe for further investigation.

## 3. Population genetics consequences of selfing

### 3.1. Theoretical expectations

Self-fertilization has drastic consequences on the patterns and distribution of genetic variation, and for the impact of natural selection. The level of selfing is therefore an important factor to consider in population genetics. Here we summarize the expected population genetic consequences of selfing (Fig. 2) and then present empirical results from *Arabidopsis* and *Capsella* that illustrate the theoretical expectations.

Selfing has two major key effects; it results in reduced heterozygosity and a reduced effective population size ($N_e$). Under complete selfing, heterozygosity is halved every generation. Hence, the heterozygosity of a completely selfing population is almost fully eliminated already after six generations of complete selfing. A side effect is a rapid generation of isolated lines of different genotypes *(40)* which is expected to result in stronger population structure in selfing species compared to outcrossers *(41)*. Moreover, selfers are expected to exhibit a reduction in $N_e$ for several reasons. First, selfing immediately results in a two-fold reduction of the number of independently sampled gametes, and this is expected to reduce the $N_e$ by a factor of two *(42, 43)*. Even greater reductions in $N_e$ are expected if selfers undergo more frequent extinction and recolonization dynamics than outcrossers *(44)*, or if the origin of selfing species is often associated with bottlenecks *(14)*. Furthermore, because selfing results in a rapid decrease in heterozygosity, recombination is less efficient at breaking up linkage disequilibrium in selfers than in outcrossers *(45)*. In this situation, background selection or recurrent hitchhiking (linked selection) will have a greater impact, reducing neutral genetic diversity beyond what would be expected in an outcrosser

*(46)*. Together, these factors decrease the overall genetic diversity *(41)* and increase the linkage disequilibrium (LD) of selfing populations *(43)*.

The combined effect of reduced $N_e$ and effective recombination rate will also affect the efficacy of selection genome-wide. On the one hand, when $N_e$ is reduced, a higher proportion of the genome behaves neutrally and alleles that were slightly deleterious in large populations become effectively neutral *(47)*. In addition, as an effect of the reduced effective recombination rate in selfers, Hill-Robertson interference *(48)* will increase and therefore limit selection efficacy further. As a consequence, one may expect selfing lineages to have an excess of non-synonymous divergence compared to synonymous divergence ($d_N/d_S$ or $K_a/K_s$) as well as polymorphisms ($\pi_N/\pi_S$), mainly due to weakened selection against slightly deleterious variants *(41)*. Further, reduced efficacy of selection and recombination rate may also decrease the level of codon usage bias *(49, 50)*. However, it should be noted that spurious signals of relaxed purifying selection can result as a result of recent demographic change *(51)*, because the time to reach equilibrium after a bottleneck is longer for nonsynonymous than for synonymous polymorphism. Ideally, forward population genetic simulations incorporating selection and demography should therefore be undertaken to validate inference of relaxed purifying selection.

The dynamics of alleles with different levels of dominance will also be affected by the mating system *(52)*, in ways that can sometimes counteract the effect of reduced $N_e$. For instance, in selfing species, increased homozygosity renders recessive alleles visible to selection, and as a result, fixation probabilities of recessive advantageous alleles are expected to be higher in selfers than in outcrossers *(53)*. Harmful recessive alleles can also be removed more efficiently, resulting in purging of recessive deleterious alleles, unless the reduction in $N_e$ in selfers is severe enough that genetic drift overpowers the homozygosity effect *(54)*.

Selfing and outcrossing populations have also been shown to differ in the dynamics of adaptive alleles. The reduced efficacy of selection will decrease the probability of slightly benefical mutation fixation but also the fixation time of benefical mutation in selfers is faster in comparison with the outcrossing species regardless of the dominance level *(13, 55)*. Further, in selfing populations, adaptation is more likely to result from new mutations (hard sweeps) while in outcrossers adaptation from standing variation (soft sweeps) is predicted to be more frequent *(13, 15)*. On the other hand, the effect of linked selection is expected to be stronger in selfing species due to reduced effective recombination rate which will increase the fixation probability of linked harmful mutations *(56)* and potentially limiting the adaptive potential of selfers.

Mating system has also been hypothesized to affect the transposable element (TE) content of the genome. TEs are mobile genetic elements that make up a large yet variable proportion of many plant genomes *(57, 58)*. Theory predicts that both mating system variation and differences in the effective population size ($N_e$) should affect TE content, because these factors affect possibilities for TEs to spread, the potential for evolution of self-regulation of transposition, and the efficacy of selection against deleterious TE insertions. For instance, outcrossing enhances opportunities for TE spread *(59)*, and transposition rates should evolve to be highest in outcrossers, which should therefore be expected to have higher TE content than highly selfing species *(60)*. Likewise, if the harmful effects of TEs are mostly recessive or codominant, increased homozygosity in selfers leads to more efficient purifying selection against TEs in selfers *(61)*. On the other hand, natural selection against slightly deleterious TE insertions could be compromised in selfing species, because of their reduced effective population size *(41)*. Increased homozygosity in selfers might also decrease the deleterious effect of TEs, because this decreases the probability of ectopic recombination *(62)*. Under this scenario, a transition to selfing would be expected to lead to an increase in TE content.

Different models therefore yield contrasting predictions regarding the expected effect of mating system variation on TE content.

**3.2 Empirical results**

Several empirical results have confirmed the theoretical predictions regarding the population genetics effects of selfing. Figure 3 summarises some empirical population genetics results from selfing and outcrossing *Arabidopsis* and *Capsella* species. First, while it is difficult to disentangle the effect of past demographic events from the effect of selfing, both *C. rubella* and *A. thaliana* present high levels of population structure *(21, 22, 63)*. Indeed, population structure is stronger in the selfer *C. rubella* than in the outcrossing *C. grandiflora*, consistent with theoretical expectations for selfers *(22)* (Fig. 3).

Furthermore, evidence for a reduction in $N_e$ has been found in natural populations of *C. rubella* and *A. thaliana.* In *C. rubella* both synonymous diversity and population recombination rate are significantly lower than in the outcrossing sister species *C. grandiflora (22, 64)*. Similarly, the selfer *A. thaliana* shows lower synonymous polymorphism in comparison with the outcrossing *A. lyrata (65, 66)* (Fig. 3). An early study also found evidence for a faster decay of linkage disequilibrium (LD) in *A. lyrata* than in *A. thaliana (66)*. However, more recent work has shown that there is also high variation in nucleotide diversity and LD patterns in *A. thaliana,* both between different parts of the genome and across different geographic regions and habitats *(67-70)*. Patterns in *A. lyrata* are also complicated by strong population size decrease in several extensively studied *A. lyrata* populations *(71)*, which also decreases the population recombination rate. Hence, it is worth noticing that both diversity and LD are both highly dependent on the past demographic history and the local variation in the recombination rate across the genome. The decay of LD also depends on whether estimates are based on local or global population samples *(72)*. In *A. thaliana*

LD decays faster in a world-wide sample in comparison with local populations *(68, 70)* which may be due to local populations having a low number of founders *(72)*.

Empirical evidence for the impact of selfing on the efficacy of selection started with early investigations of divergence and polymorphism in *A. thaliana* and its outcrossing sister species *A. lyrata*, using a limited number of loci. This study found very limited evidence for relaxed selection in *A. thaliana* *(73)*. However, using genome-wide polymorphism data, Slotte et al. (2010) *(74)* found evidence for weaker purifying selection on nonsynonymous sites in *A. thaliana* relative to the outcrosser *C. grandiflora*. Further studies confirmed decreased codon usage bias in both *A. thaliana* and *C. rubella* in comparison with the outcrossing *A. lyrata* and *C. grandiflora* *(50)* (Fig. 3E). Analyses of population genomic data from *C. rubella* and *C. grandiflora* further found evidence for a higher ratio of non-synonymous to synonymous polymorphism in *C. rubella* *(75, 76)*. Forward population genomic simulations demonstrated that this was likely primarily a result of the reduced $N_e$ in *C. rubella*, and not due to a major shift in the distribution of fitness effects (DFE) in association with the shift to selfing *(76)*. More recently, a study that directly estimated the DFE based on analyses of site frequency spectra found evidence for a higher proportion of nearly neutral nonsynonymous mutations in the selfers *C. orientalis* and *C. bursa-pastoris* relative to the outcrosser *C. grandiflora* *(18)*. These results are in general agreement with the theoretical expectation that purifying selection should be relaxed in selfers.

While some models predict that a shift to selfing should lead to a reduced prevalence of TEs, other models predict the opposite. So far, empirical evidence from *Arabidopsis* and *Capsella* do not unequivocally support either of these predictions. On one hand, the comparison of structure of the *A. thaliana* and *A. lyrata* genome sequences suggested that the *A. thaliana* genome contained large numbers of small deletions, especially in TEs *(77)*. This result is consistent with the hypothesis that in selfing species TEs are more efficiently removed *(61)*. One the other hand, Lockton & Gaut (2010) *(78)* found that many TE families are in higher frequency and are subjected to weaker selection in *A. thaliana* in comparison with *A. lyrata*. Furthermore, comparison of *C. rubella*, *C. grandiflora*, *A. thaliana* and *A. lyrata* revealed that the TE frequency and density in *Capsella* showed a stronger resemblance to *A. thaliana* than to *A. lyrata* *(76)*. This may indicate that the reason for the TE abundance difference between the *Arabidopsis* species is accumulation of TEs in the *A. lyrata* genome rather than decline in the selfing *A. thaliana* lineage.

Focusing on TE content in selfing and outcrossing *Capsella* species, Ågren et al. (2014, 2016) *(79, 80)* found an increase in TE number in *C. rubella* but a slight decrease in the selfer *C. orientalis*, in comparison with the outcrossing *C. grandiflora*. In the polyploid selfer *C. bursa-pastoris*, no evidence for a difference in TE dynamics was found in comparison with its parental species, *C. grandiflora* and *C. orientalis* *(80)*. Thus, while there is some evidence for a reduced prevalence of TEs in selfers, the results are not unequivocal, and further work is needed to clarify whether the contrasting findings might be related to the timing of the shift to selfing and demographic history. Indeed, there is evidence for an effect of demographic history on selection against TEs in *A. lyrata*, where large refugial populations exhibited a signature of purifying selection against TE insertions, whereas in bottlenecked populations, TEs were evolving neutrally *(81)*. In addition to broad comparative genomic studies contrasting species that differ in their mating system, studies of intraspecific variation can thus provide insight into population genomics and selection on TEs *(82)*. Because TEs are important contributors to variation in plant genome size and TE silencing can also affect gene regulation *(83-85)*, it is of considerable interest to improve our understanding of the impact of mating system on variation in TE content.

**4. Discovering the geographic origin and the timing of the mating system shift**

Understanding the timing, mode, and geographic location of the shift to selfing is of key importance for proper interpretation of population genomic data from selfing species *(86)*. For instance, improved understanding of the timing and geographical location of the shift can be key for interpretation of genetic structure, and can allow one to account for underlying neutral (demography induced) processes when investigating changes in the efficacy of selection.

In *A. thaliana*, several studies have estimated the timing of the emergence of selfing based on patterns of polymorphism and demographic modeling. *A. thaliana* is native to the Eurasia and Africa *(87-89)* and widely spread especially in Europe. It has also recently spread into North America in association with humans *(21)*. All the currently known accessions are SC, indicating that the evolution of this trait preceded the worldwide spread of the species. Linkage disequilibrium patterns suggested that the transition to selfing in *A. thaliana* occurred as early as 1 000 000 years ago *(90)* while coalescent modeling using *S*-locus diversity suggested a younger origin with an upper estimate of approximately 400 000 years ago *(91)*.

The recent development of large-scale population genomics datasets from 1135 *A. thaliana* accessions *(63)*, offers one of the best population genomics resource for studying plant population genetics and molecular evolution. A recent demographic modeling study exploited this resource and included additional accessions covering roughly the African distribution of the species to investigate the timing and geographic origin of the shift to selfing in *A. thaliana (89)*. Using a combination of the MSMC method that infers cross-coalescent times and fluctuations in the effective population size using a whole-genome data from multiple populations *(92)* and the site frequency spectrum based diffusion approximation method δαδi *(93)*, Durvasula et al. (2017) *(89)* inferred the demographic history of the different groups from Africa and Europe. They suggest that partial loss of SI occurred 500 000-1 000 000 years ago subsequent to the migration of the ancestral founding *A. thaliana* population to Africa approximately 800 000-1 200 000 years ago. Although the existence of multiple non-functional *S*-haplotypes suggests that the final loss of SI occurred multiple times independently *(94, 95)*, the new study including African accessions shows that all the currently known *S*-haplotypes are found coexisting in Morocco, suggesting that selfing originated in this geographic region *(89)*. This result was further supported by the higher estimated $N_e$ in African populations, and they estimated that the species spread out-of-Africa some 90 000-140 000 years ago.

The post-glacial colonization of *A. thaliana* within Europe has likely proceeded through acquirement of a weedy lifestyle *(63)*. The first European colonists likely occurred in the southern and eastern part of Europe *(63, 96, 97)* and the massive spread over central and northern parts of Europe occurred later, possibly associated with human action *(63)*. Population structure analysis revealed that there are two distinct groups in Central Europe and these two groups have admixed in Central Europe *(98-100)*. Two distinct lineages are also present in Scandinavia with accessions from Finland as well as from the northern parts of Sweden and Norway forming their own cluster while the southern Swedish accessions cluster with the southern accessions *(99, 101)*. Using a whole-genome population-genomics approach Lee et al. (2017) *(102)* found five different clusters within European *A. thaliana* accessions and they suggest that that these groups have given rise to the current distribution of the species within Europe.

Another selfing example from *Arabidopsis*, where the demographic and colonization history has been studied, is the polyploid species *A. suecica*, which is a selfing allopolyploid between *A. thaliana* and *A. arenosa* (Fig. 1). The species is spread in central Sweden and southern Finland. Early investigation explored 52 microsatellites and four nuclear sequences *(103)* and inferred a single and recent origin of the species approximately 12 000-300 000 YA followed by northward spread using a Bayesian coalescent population modeling. This single origin has also been supported by other studies based on the amount of variation in chloroplast sequence data *(104, 105)*. However, recent investigation of whole-genome re-sequencing data from 15 *A. suecica* accessions concluded that the multiple origins hypothesis cannot be ruled out *(106)*. Based on the *S*-locus haplotype dominance patterns in these accessions they suggest that *A. suecica* could have been SC, at least to some degree, immediately after the species emergence approximately 15 100-16 600 years ago (Fig. 1).

In *Capsella* it has been estimated that the timing of the loss of SI is much more recent. Isolation-migration analyses based on 39 gene fragments and assuming a mutation rate of $1.5 * 10^{-8}$ suggested that the shift to selfing was concomitant with speciation of *C. rubella* from an outcrossing ancestor similar to present-day *C. grandiflora,* and that this occurred some 20 000 years ago *(64)*. Likewise, an investigation of the diversity patterns at the *S*-locus across the European range of the species suggested that loss of SI took place in Greece approximately 40 000 years ago *(37)*, since the presumably ancestral long form of *SRK* allele is present in this region while the other accessions harbor a shorter form. Divergence estimates based on analysis of genome-wide founding haplotypes suggested that selfing evolved later, approximately 50 000 to 100 000 years ago *(75)*. Assuming a different mutation rate of $7.1 * 10^{-9}$, Slotte et al. 2013 *(76)* analyzed genome-wide site frequency spectra using δaδi, estimated that the timing of the split between *C. grandiflora* and *C. rubella* occurred <200 000 years ago (Fig. 1). Later on, coalescent-based analyses of genome-wide joint site frequency spectra from *C. grandiflora, C. orientalis* and *C. bursa-pastoris* were used to infer the timing of allopolyploid speciation resulting in the selfing species *C. bursa-pastoris* (*(18)*, Fig. 1). Together these analyses have provided an evolutionary framework for further genomic studies of the consequences of selfing and allopolyploidy in *Capsella* (Fig. 1).

**5. Some caveats**

As many authors point out, estimates of population split times are usually based on assumptions that contain considerable uncertainty (see e.g. *(103)*). For example, a fixed mutation rate or a distribution around a mean is often assumed. The direct mutation rate estimate from *A. thaliana* mutation accumulation lines of $7.1 * 10^{-9}$ *(107)* is commonly used for *Arabidopsis*, whereas, as pointed out above, early estimates in *Capsella* were based on a mutation rate of $1.5 * 10^{-8}$ *(108)* while later studies, e.g. Slotte et al. (2013) *(76)* and Douglas et al. (2015) *(18)* have used the Ossowski et al. (2010) *(107)* mutation rate estimate. Other assumptions may include constant generation time, recombination rate and analyses may further be restricted to a limited number of demographic models. These factors may be variable between studies and it is important to pay attention to these details when evaluating the results of demographic modeling.

A more severe limitation concerns the effect of selection on linked sites on demographic inference. Simulations have shown that the increased intensity of background selection in selfers can rapidly lead to a strong reduction in neutral diversity *(109)*. Based on these findings, some authors have questioned whether it is possible to reliably infer demographic changes associated with the shift to selfing based on neutral polymorphism *(109)*. The impact of linked selection on patterns of neutral polymorphism is indeed a general problem for demographic inference *(110-112)*. To some degree, it may be possible to circumvent this problem by judiciously choosing which sites to use for demographic inference *(46)*, and by using forward population genetic simulations in software such as SLiM *(113, 114)* to assess whether results are robust to the effect of selection on linked sites. As an example, a recent study on *Arabis alpina* used site frequency spectra for sites in genomic regions with high recombination rates and low gene density, which should be least affected by linked selection, to infer the demographic history of selfing Scandinavian populations *(115)*. The reliability of this inference was further checked with forward simulations incorporating background selection and a shift to selfing *(115)*. This study therefore demonstrated one way to assess the effect of linked selection on demographic inference in selfing populations.

**6. Future directions**

In this chapter, we have given a brief overview of population genomic studies of the transition to selfing, focusing on the two model systems *Arabidopsis* and *Capsella*. One limitation of most of the studies presented here is that they have focused on comparing pairs of species with contrasting mating systems. Ideally, to be able to distinguish between idiosyncracies of one particular contrast and effects of selfing *per se*, future comparative population genomic studies of the effect of selfing should include multiple phylogenetically independent contrasts.

An additional limitation is that most studies have focused on contrasting only highly selfing and obligate outcrossing species, although there is a lot more diversity to plant mating systems. For instance, a substantial proportion (approximately 42%) of flowering plants undergoes a mix of outcrossing and self-fertilization *(116)*. Despite this fact, and despite the existence of theoretical and simulation-based work on the expected population genomic consequences of partial selfing *(15, 54, 56, 117)*, there is a dearth of empirical population genomic studies including mixed mating populations and species (but see e.g. *(115, 118)*). One exception is a recent study which tested for a difference in the impact of purifying selection among obligate outcrossing, mixed-mating and highly selfing populations of *Arabis alpina (115)*. This study found no major detectable difference in purifying selection between mixed mating and outcrossing populations, wheras purifying selection efficacy was significantly lower in Scandinavian selfing populations, most likely as a result of a postglacial colonization bottleneck *(115)*. These results are consistent with the expectation that a low level of outcrossing may be sufficient to prevent accumulation of deleterious alleles *(117)*. However, further empirical studies in more mixed mating species and populations, ideally including larger sample sizes, are required to establish the generality of this pattern.

Another less empirically studied issue is how the prevalence of positive selection and especially how selective sweeps are impacted by mating system. In *A. thaliana*, the proportion of amino acid substitutions driven by positive selection has been estimated to be close to 0% *(119, 120)* while for example in the outcrossing relative *C. grandiflora* this proportion has been estimated to be as high 40% *(119)*. However, making inferences on the underlying cause of this difference is not straightforward. For example, positive selection has also been shown to be rare in the outcrossing *A. lyrata (121)* suggesting that other factors, such as for instance demographic history, may also result in low rates of adaptive substitutions. The theoretical work on the effect of dominance on fixation probabilities has yielded results that could be tested by taking advantage of genome-wide population genetics datasets and methodological developments regarding sweep detection, e.g. *(122, 123)*. For example, with such methodology, it is possible to test whether there is a difference in the relative occurrence of hard and soft sweeps in selfers and outcrossing populations as predicted by theoretical work *(13)*. Further advances in the classification of mutations into different dominance classes *(124)* will allow testing hypotheses related to the behavior of recessive, additive, and dominant alleles in selfing vs. outcrossing species.

**Figure legends**

Figure 1. Schematic drawing of evolutionary relationships among the most well-studied *Arabidopsis* (A) and *Capsella* species (B). Mating system (selfing or outcrossing) and self-incompatibility status (SI: self-incompatible; SC: self-compatible) and ploidy level is indicated for each species. Approximate estimates of split times are indicated by arrows. For *A. thaliana* and *A. lyrata*, we show two recent estimates based on Guo et al. (2017) *(125)* and Beilstein et al (2010) *(126)*. The estimate of the timing of the origin of *A. suecica* is based on Novikova et al. (2017) *(106)*. The timing of the population split between *C. rubella* and *C. grandiflora* is based on Slotte et al. (2013) *(76)* and the timing of the origin of *C. bursa-pastoris* and the split between *C. orientalis* and the *C. grandiflora*/*C. rubella* lineage is based on Douglas et al. (2015) *(18)*. Photographs of Arabidopsis species were taken by Jon Ågren (*A. thaliana*), Robin Burns (*A. suecica*), Johanna Leppälä (*A. arenosa*), Tiina Mattila (*A. lyrata*), Vincent Castric (*A. halleri*), and Rie Shimizu-Inatsugi (*A. kamchatica*). All *Capsella* photographs were taken by Kim Steige.

Figure 2. A network showing the effect of selfing on population parameters (orange) and population genetics statistics (pink). The box shape indicates the predicted effect (increase/decrease) of selfing on each factor (square boxes indicate that the effect can be either an increase or a decrease, depending on the exact combination of parameter values).

Figure 3. Empirical results illustrating the impact of a mating system shift in the *Capsella* and *Arabidopsis* genus. **A. Population structure:** Global $F_{ST}$ among populations is higher in the selfer *C. rubella* than in the outcrosser *C. grandiflora (22)*. Elevated value of $F_{ST}$ are also found genome-wise among populations of *Arabidopsis thaliana (70)*. **B. Neutral genetic diversity:** nucleotide diversity at synonymous sites is lower in selfers compared to their outcrossing relatives in both *Capsella* and *Arabidopsis (76, 119, 127)*. **C. Strength of purifying selection:** a higher ratio of non-

synonymous to synonymous nucleotide diversity suggests relaxed purifying selection in the selfing *C. rubella (76, 119)*. In contrast, in *Arabidopsis,* the outcrosser *A. lyrata* has a higher ratio of non-synonymous to synonymous nucleotide diversity than *A. thaliana*. **D. Distribution of fitness effect** (DFE) in bins of $N_e s$ (the product of the effective population size and the selection coefficient) for new nonsynonymous mutations *(18, 83, 119, 127)*. **E. Codon usage bias:** maximum likelihood estimates of selection coefficient, ϒ, for the ten amino acids with 2-fold degenerate codons between the selfing and outcrossing species. Whiskers are 95% confidence intervals obtained by the MCMC analysis (see Qiu et al. 2011 *(50)* for details, with permission from GBE).

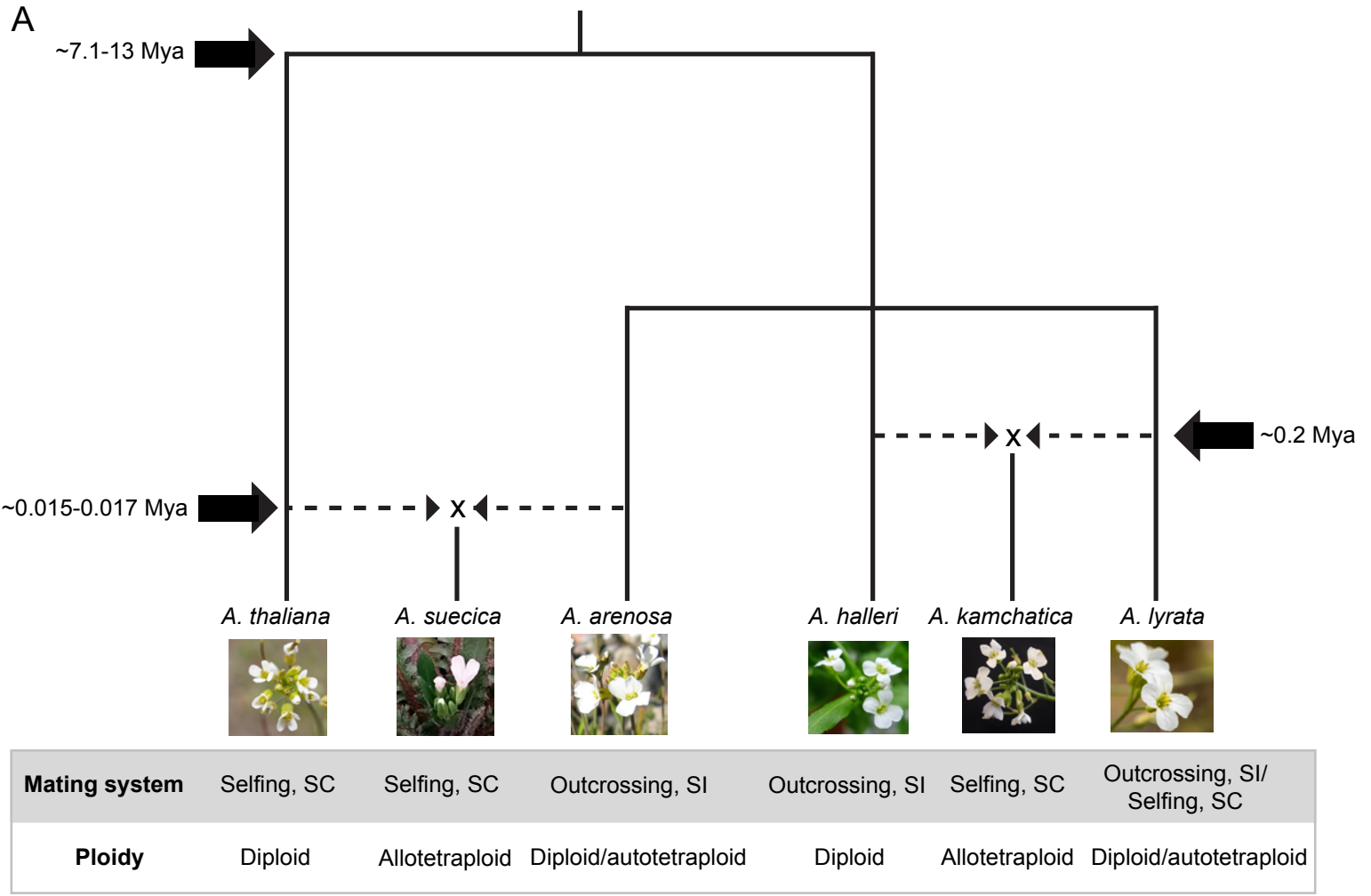

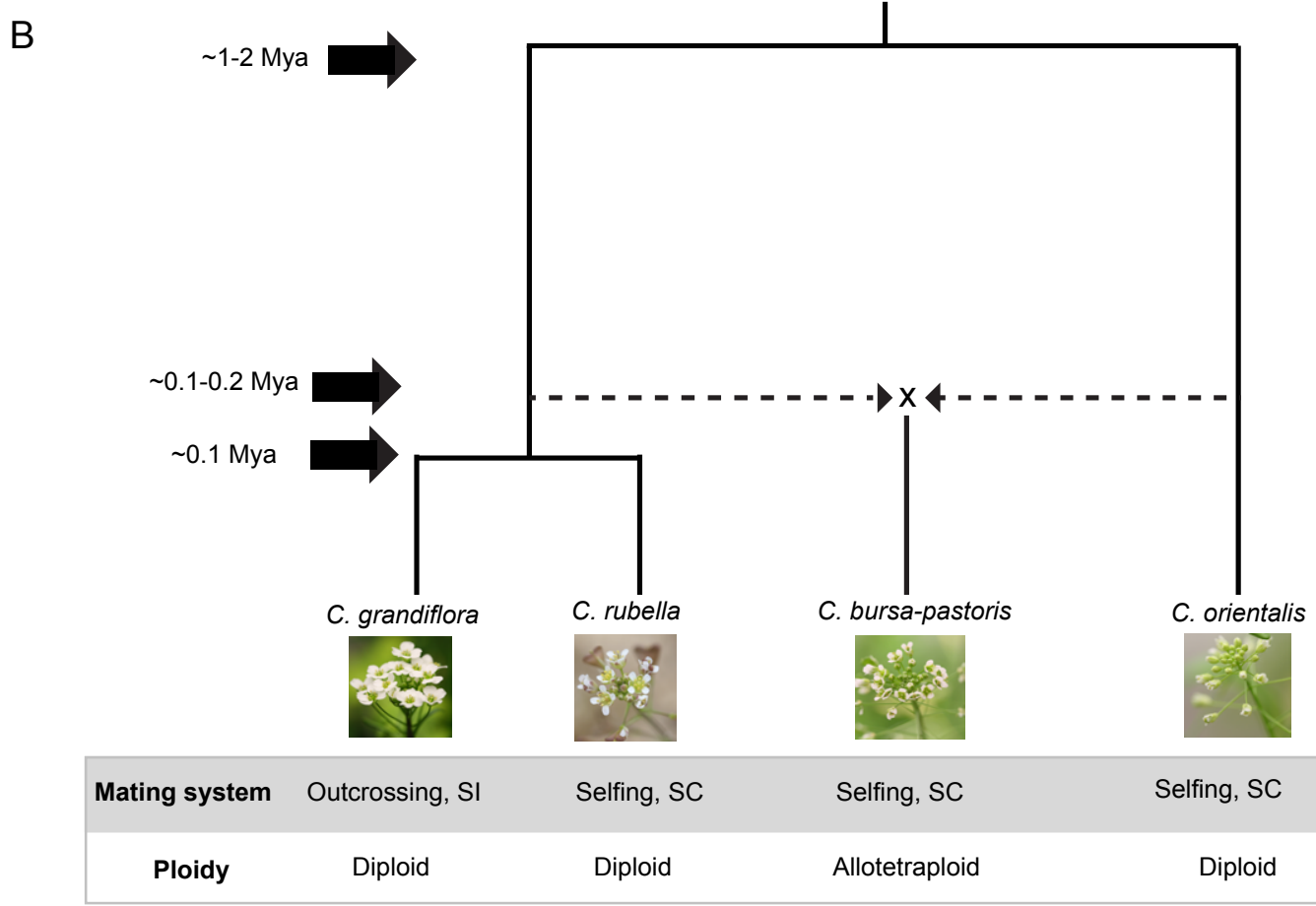

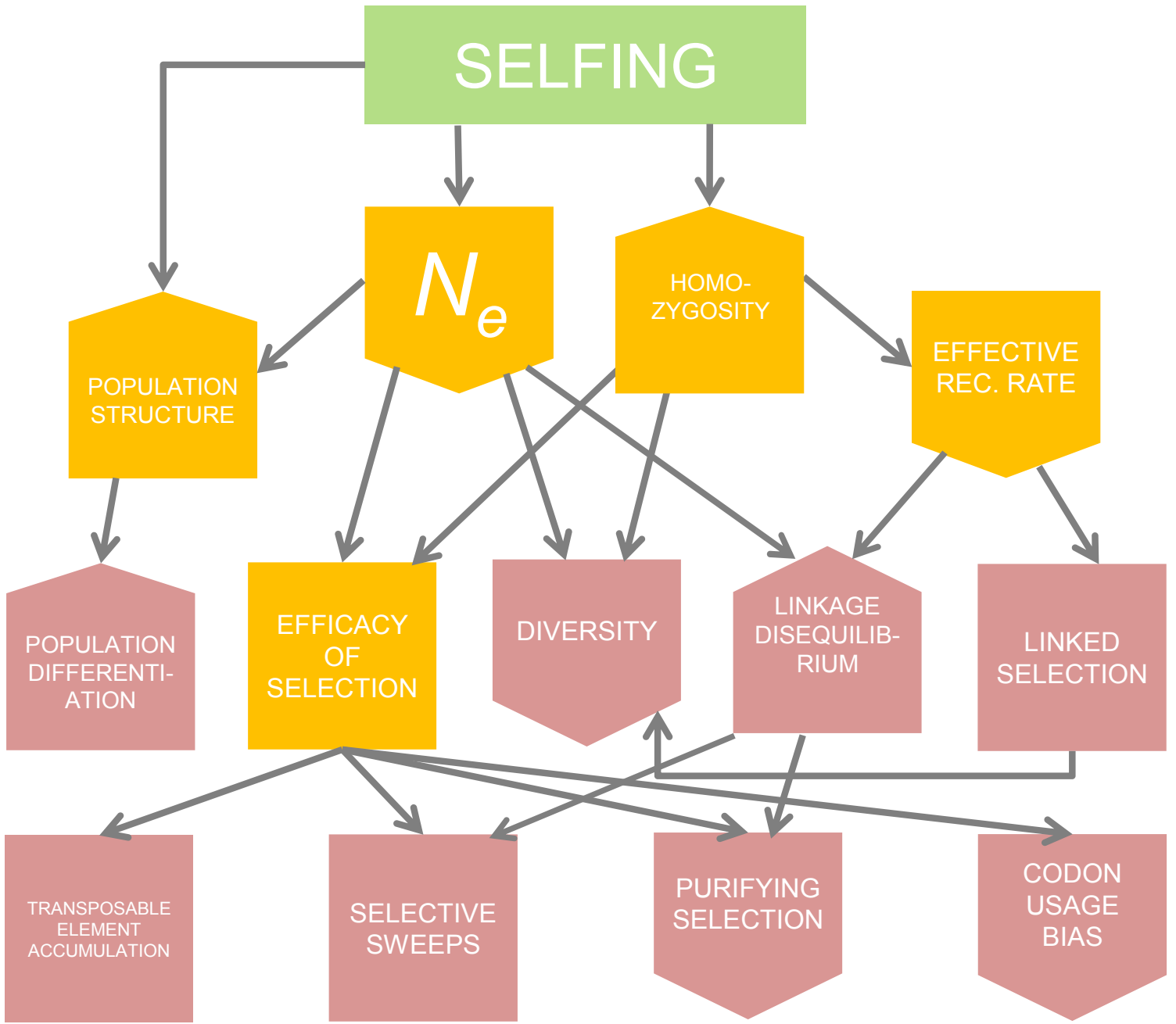

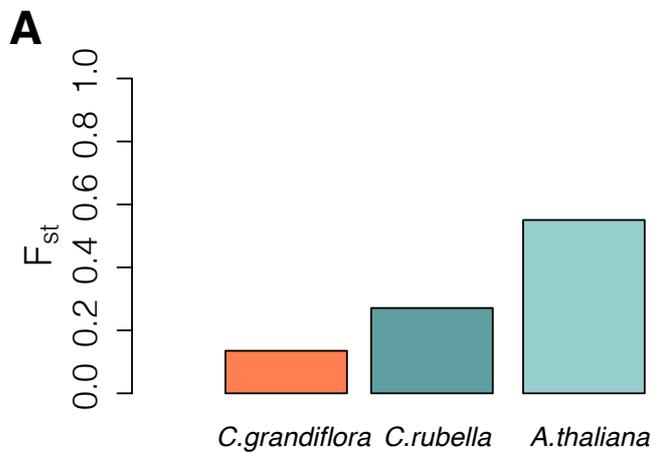
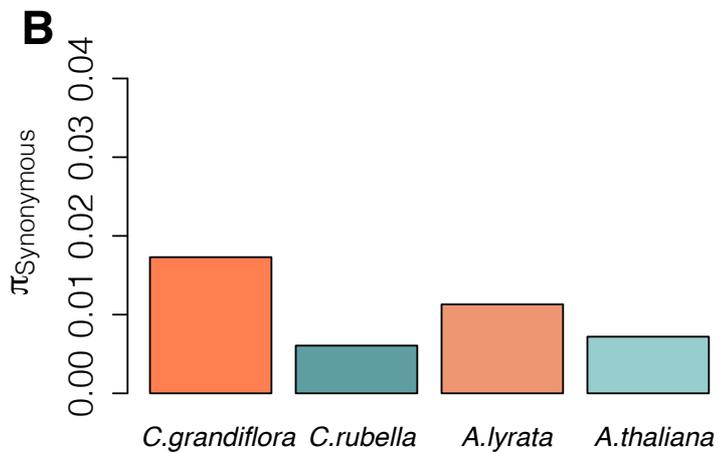
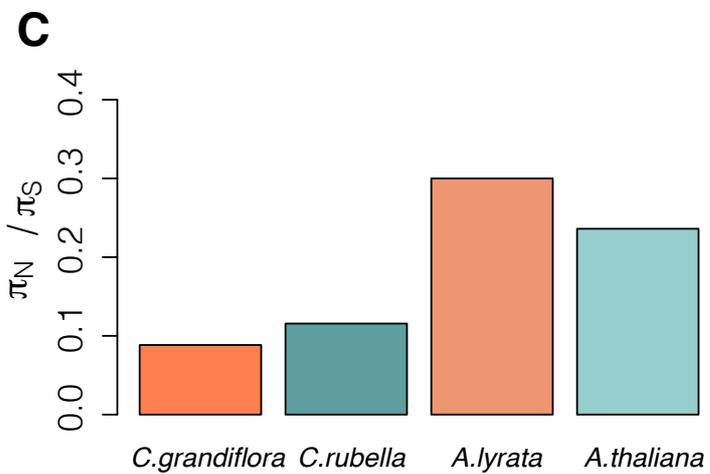
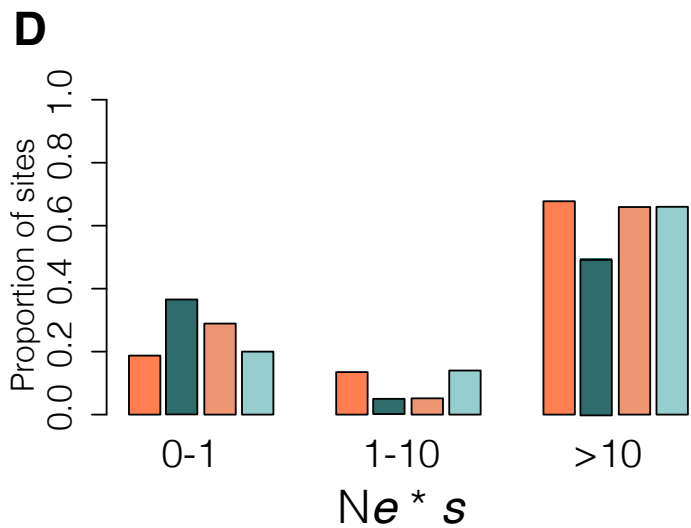
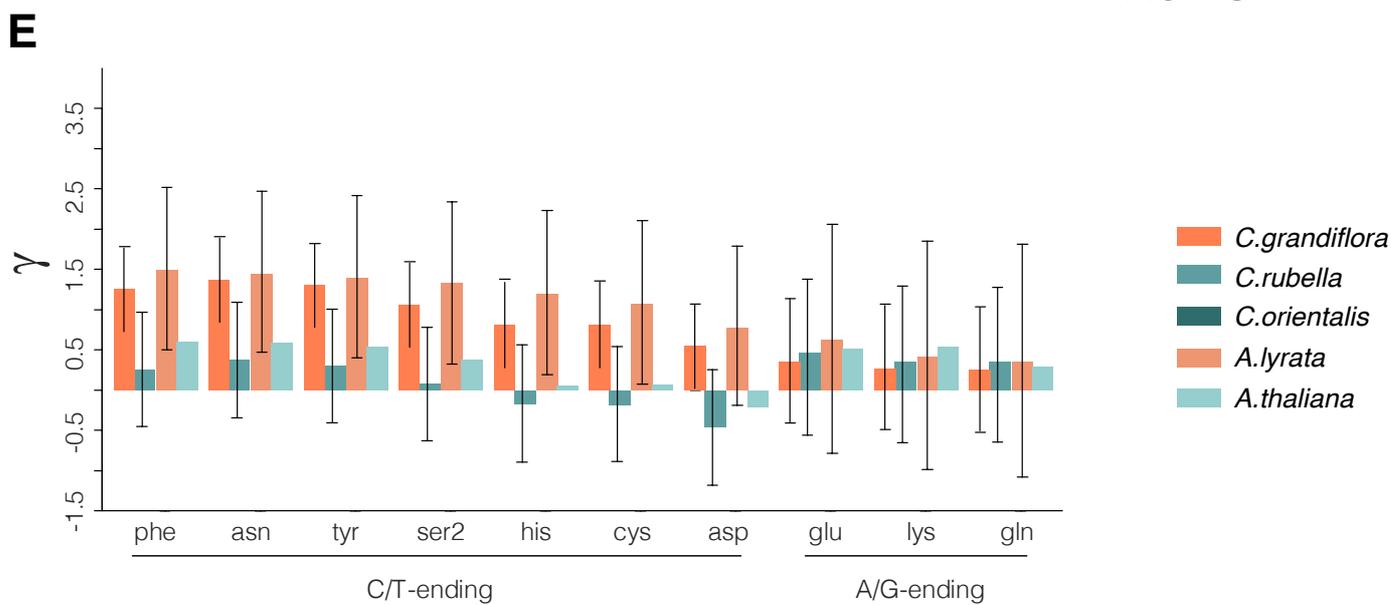